\documentclass[aps,english,prb,reprint,superscriptaddress, citeautoscript,cite]{revtex4-1}

\usepackage{graphicx}
\usepackage{amssymb}
\usepackage{babel}
\usepackage{color}
\usepackage{multirow}
\usepackage{times,mathptmx}
\usepackage{epstopdf}
\usepackage[version=3]{mhchem}

\begin{document}

\title{Interaction of Adatoms with Two-Dimensional Metal Monochalcogenides (GaS, GaSe)}

\author{H. D. Ozaydin}
\affiliation{Department of Physics, Adnan Menderes University, Aydin 09010, Turkey}

\author{Y. Kadioglu}
\affiliation{Department of Physics, Adnan Menderes University, Aydin 09010, Turkey}

\author{F. Ersan}
\affiliation{Department of Physics, Adnan Menderes University, Aydin 09010, Turkey}

\author{O. \"{U}zengi Akt\"{u}rk}
\affiliation{Department of Electrical and Electronic Engineering, Adnan Menderes University, 09100 Ayd{\i}n, Turkey}

\author{E. Akt\"{u}rk} \email{ethem.akturk@adu.edu.tr}
\affiliation{Nanotechnology Application and Research Center, Adnan Menderes University, Ayd{\i}n 09010, Turkey}

\date{\today}

\begin{abstract}
In this paper, we report first principles calculations based on density functional theory to reveal the effects of selected adatoms (Li, Na, K, Be, Mg, Ca, B, C, N, O, Al, Si, P, Ga, Ge, As, Se and S) adsorption on GaX ( where X=S, Se) monolayers.  
It is found that all adatoms adsorbed on  GaX monolayers can form strong chemisorption bonds except for Mg atom due to weak bonding nature of Mg atom with early 3d transition metals.  Most of the adatoms of the same group elements of the periodic table are bound to GaX substrates at the same adsorption sites. All adatoms stayed above the upper layer of GaX sheets except for Be adatom on GaSe substrate and overall hexagonal geometry has been retained.  The electronic structures are modified by locally through the supercell calculations. Specific adatoms, such as C, Si, Ge, N, P and As give rise to spin polarization and attain integer magnetic moments and hence can contribute half metallic character to the system.  Our results can serve as a basis for future experimental and theoretical studies of adsorption on GaX.  

\end{abstract}

\maketitle
\section{Introduction}

Recently, monolayer structure of metal monochalcogenides  have been produced with exfoliation techniques and   the electronic, magnetic, mechanical and optical 
properties of these monolayers\cite{yuan2016arrayed,li2014controlled,mahjouri2014digital,xu2013graphene,hu2012synthesis,yagmurcukardes2016mechanical} have been an active subject of theoretical and experimental studies. These are GaX, in which there are two sheets of Ga layers sandwiched between chalcogenide layers (in order of X-Ga-Ga-X, where X=S, Se, Te). Even before most research has focused on monolayer structures of GaX, bulk GaX with layered structure have been  widely used in diverse fields due to many interesting electronic and optical properties. \cite{leontie,shi,allak,gasanly,yuksek}
These layered structures have a highly anisotropic bonding force, because layer-layer interaction has very weak van der Waals (vdW) force than the bonding force within the layer. GaX is consisted of both covalent Ga-Ga bonds and cation(Ga)-anion(X) bonds by contrast with many other layer compounds, and due to this extremely weak vdW force they can be easily cleaved along layers. There have been a number of experimental efforts to fabricate the two-dimensional (2D) GaX monolayers\cite{late2012gas,late2012rapid,lei2013synthesis} since they are stable layered 
semiconducting materials with wide band gaps. Based upon only comprising light elements of GaX monolayers, they have attracted interest due to their potential applications in fields such as solar energy conversion\cite{levy1992photoelectrochemistry}. 
For instance, multilayer GaS$_{1-x}$Se$_x$ (0$\leq$x$\leq$1) were 
synthesized by means of chemical vapor transport (CVT) by Jung \textit{et al.} and showed that GaS-GaSe alloys have band gaps in the range of 2.0-2.5 eV which are matching 
the visible region of spectrum with red to green\cite{jung2015red}.  Tongay \textit{et al.} have recently reported that the effects of different gas molecules on photoelectric 
response of few-layer defected GaSe phototransistors\cite{yang2016highly}, and they proved that existing of O$_2$ molecule can increase the performance of GaSe phototransistors compared
to that in the air. However, similar study on GaS monolayer showed that photo-response of GaS nanosheet is higher in NH$_3$ environment instead of that in the air or O$_2$ 
environment\cite{yang2014high}. 

In fact, GaSe triangular monolayer flakes are grown on SiO$_2$/Si substrates, but these flakes show randomly distributed orientations, therefore, when
these triangulars merge to form larger sheet, grain boundaries occur and that cause to decrease electrical properties\cite{li2014controlled}. Recently, the same group synthesized
preferred orientations of 2D GaSe atomic layers on graphene by vdW epitaxy, therefore larger GaSe monolayers can grow\cite{li2015van}. At the same time, dielectric and electronic properties of monolayer and multilayer GaS and GaSe are investigated and obtained that electron energy low-loss spectra shifts towards larger wavelength with 
decrease of layer thickness\cite{li2015ab,ma2013tunable}. Furthermore, Peng $\textit et$ $\textit al.$ systematically increased Mg doping in GaS and GaSe monolayer which causes p-type carriers in GaSe nanosheets\cite{peng2014characteristics}. While GaS and GaSe monolayer is a semicondutor, Ga vacancy induces ferromagnetism and half-metallicity, however S or Se hole doping rendering GaS and GaSe \textit{n}-type semiconductors\cite{wu2014magnetisms,cao2015tunable,ao2015functionalization}. Moreover, substitutional impurity effect is studied for GaX (X=S, Se) monolayers by doping transition metal and nonmetal atoms where the results give a net magnetic moments increase from 1 to  5 $\mu_B$\cite{ao2015functionalization,chen2015influential}. 

Although there are numerous experimental and theoretical studies on the exfoliation of GaX monolayers or the investigation of their structural, optical, and electronic properties, 
we have not encountered any studies about the effects of adatom adsorption on GaS or GaSe monolayer (except of gas molecule adsorption\cite{jung2015red,yang2016highly,yang2014high}). 
In contrast to MX$_2$ or MX$_3$ monolayers, GaS and GaSe monolayers include light elements only, so it will be important to investigate how the electronic and 
magnetic properties of GaS (or GaSe) monolayers change when the light elements adsorbed on them. For this reason, in this study, we systematically investigate the effects of 
light elements (Li, Na, K, Be, Mg, Ca, B, C, N, O, Al, Si, P, Ga, Ge, As, Se and S) on the GaS and GaSe monolayers.  Our study reveals interesting results, which are important for further study and applications of GaX monolayers. These are as follows: (i) Our results indicate that all adatoms except Mg are bound to  GaX monolayers with significant binding energy. (ii) Mostly the same group elements of periodic table, prefer to locate at the same adsorption sites for both on GaX substrates.  The electronic structures are modified by adatom adsorption and the nonmagnetic indirect band gap semiconductors attain net magnetic moment  through the adsorption  C, Si, Ge, N, P and 
As  adatoms.

\section{Computational Methodology}\label{comp}
We have performed  first-principles plane wave calculations based on the spin-polarized density functional theory (DFT)\cite{Kohn-Sham} using projector augmented wave (PAW)\cite{blochl1994projector} potentials. The exchange correlation potential is 	approximated by a generalized gradient approximation (GGA) using the Perdew-Burke-Ernzerhof (PBE) functional \cite{perdew1996generalized} . The van der Waals (vdW) correction to the functional is included by using the DFT-D2 method of Grimme\cite{grimme2006semiempirical}.  Adsorption of all adatoms  are investigated  using GaX supercells. Electronic and geometric relaxations of the structures are performed using $(4 \times 4)$ 
supercells for GaS (or Se) sheets which are large enough to avoid 
interactions 
between adjacent adatoms, since, this size of cell is 4-5 times larger than dimer length of considered adatoms. The vacuum spacing between the layers are taken at least 15 \AA. 
Based on the convergence analysis on kinetic energy cutoff and \textbf{k}-point 
sampling, a plane-wave basis set with kinetic energy cutoff is taken to be $\hbar^2\arrowvert\textbf{k+G}\lvert^2/2m=$ 400 eV (600 eV for oxygen adsorption) and 
the Brillouin zone (BZ) is sampled in the \textbf{k}-space within $(4 \times 4 \times 1)$ $\Gamma$-centered. Atomic positions are optimized 
using the conjugate gradient method, where all the atomic coordinates are fully relaxed until the force is 
less than 0.001 eV/\AA~ and the energy convergence value between two consecutive steps is chosen 
as $10^{-5}$ eV. Gaussian type Fermi-level smearing method is used with a smearing width of 0.01 eV. 
The total energies and electronic density of states of bare and adatom adsorbed structures are calculated by using PBE method. Nevertheless, energy band gaps 
are underestimated by DFT, we corrected bare GaS and GaSe monolayer band energies by using Hybrid Heyd-Scuseria-Ernzerhof (HSE) 
method\cite{HSE,heyd2006erratum,paier2006screened}.  All the numerical calculations are performed using the Vienna \textit{ab-initio} Simulation Package (VASP)\cite{vasp3,vasp4} code. 

The magnetic ground states of adatom+substrate systems revealed by the optimization based on spin-polarized DFT are further elaborated. In order to obtain the magnetic 
moments, the structure is first completely relaxed by spin-polarized calculations. Charge transfer between adatom and 
monolayer GaS/GaSe $\Delta \rho$, is calculated by using Bader analysis\cite{Henkelman}. The difference charge density, $\Delta \rho = \rho [A+GaX]-\rho [GaX]-\rho[A]$, 
is calculated by subtracting the charge density of pristene GaS/GaSe sheets $\rho[GaX]$, and that of free adatom $\rho[A]$, from the total charge density of the optimized 
adatom+substrate structure $\rho[A+GaX]$. Positive values of $\Delta \rho$ correspond to charge donation from the adatom to substrate, negative sign for vice versa. 
The adsorption energies of adatoms are obtained by E$_{ads}$ = E$_{adatom}$~-~E$_{GaX}$~-~E$_{adatom+GaX}$. 
Here E$_{adatom}$, E$_{GaX}$, E$_{adatom+GaX}$ are 
the ground state energies of isolated adatom, bare GaS/GaSe sheets, and the adatom adsorbed GaS/GaSe system, respectively.

\begin{figure}
\includegraphics[scale=0.40]{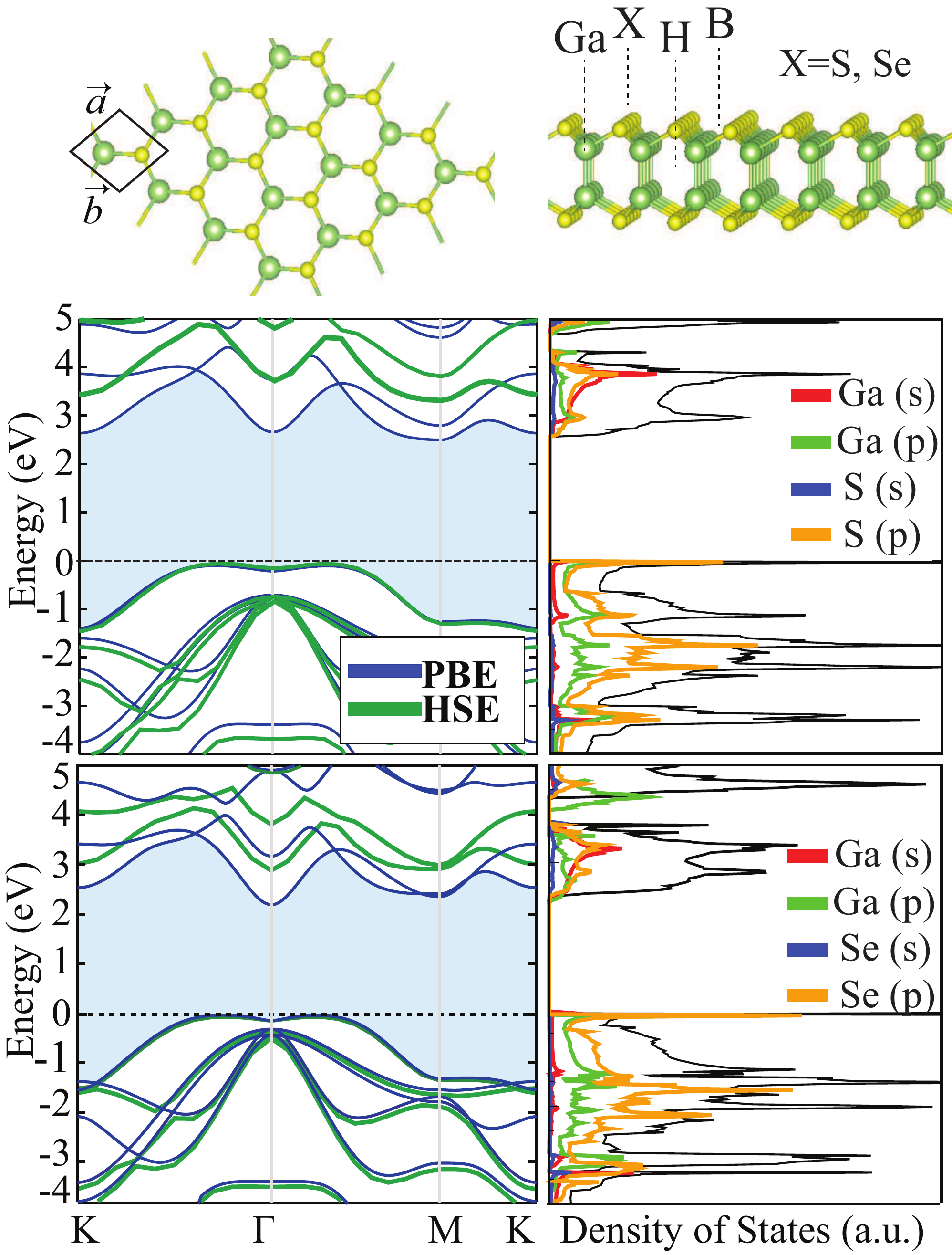}
\caption{(Color online) Top and side views of the atomic configuration of the $(4 \times 4)$ supercell GaX used to treat the adsorbed single adatom. 2D hexagonal primitive unitcell 
is delineated by dashed lines. Possible adsorption sites: the hollow (H) site on the center of a hexagon, the bridge (B) site on the midpoint of a Ga-X bond, the top (Ga) site
directly above a gallium atom, and the top (X) site directly above upper layer chalcogen atom (S or Se). Electronic band structures of the bare single layer GaX and the total
and orbital projected densities of states (PDOS). Electronic energy bands calculated by both PBE and HSE corrections.}
\label{f1}
\end{figure}
\section{Bare GaX (X~=~S, Se) Monolayer} 

Before discussing the adatom adsorption, we first examine the structural and electronic properties of bare 2D hexagonal GaX monolayer structures for understanding and analysis of interaction between bare system with adatoms. Atomic configuration of GaX monolayer is shown in Fig.~\ref{f1}, where the crystal consists of 4 atoms (X-Ga-Ga-X) in unitcell 
and X (chalcogenide atom) stands for S or Se atoms. 

The  calculated  values of lattice constants for GaS and GaSe monolayer in equilibrium  are 3.58 \AA~ and 3.75 \AA~, respectively. The bond distance between Ga and X atoms is found as d$_{Ga-S}$= 2.35 \AA~ and d$_{Ga-Se}$= 2.47 \AA~ for Ga-Se. Thickness values of 4.67 \AA~ and 4.83 \AA~ are calculated for GaS and GaSe monolayers and all these results are in fair agreement with  the previous study \cite{yagmurcukardes2016mechanical}. 

It is well known that electronic properties are affected by structural properties. The difference in chalcogen atoms of S and Se, have a little influence on the electronic properties as discussed below. GaS and GaSe monolayer structures are both semiconductors with indirect band gap as shown in Fig.~\ref{f1}. The present PBE calculations predict that monolayer GaS has a band gap of 2.59 eV which
the maximum of valance band occurring between K and $\Gamma$ points and the minimum of the conduction band at M point. However, including HSE correction increases this band gap  to 3.50 eV which is in good agreement with the theoretical value of 3.40 eV previously reported. 
Band gap values of GaSe monolayer with PBE and HSE corrections are 2.23 eV and 3.15 eV respectively. The valance band maximum (VBM) is lying between K and $\Gamma$ points similar with GaS monolayer, but the conduction band minimum (CBM) resides at $\Gamma$ point. The VBM energy level descends at $\Gamma$ point occurring two symmetric points nearby $\Gamma$ point , therefore forming the Mexican hat shape\cite{li2015ab} both in GaS and
GaSe band structures.

Density of states (DOS) of bare GaS and GaSe monolayers are displayed in Fig.~\ref{f1}, which is found that the electronic states near the Fermi level are contributed mainly
from S-3p (Se-4p), Ga-4s and Ga-(3p) states. While VBM near the Fermi level (E$_F$) are predominantly derived from the S-3p (Se-4p) states, small contributions come from Ga-3p states; CBM originate from mostly S-3p (Se-4p) and Ga-4s state hybridization.

Bader charge analysis indicates that about 0.8 electrons are transferred from Ga atoms to S atom while the amount of charge depletion from Ga to Se atoms is 0.6 electrons 
which can indicate an ionic type bonding between Ga and X atoms.

\section{Adsorption of Adatoms to GaX (X~=~S, Se) Substrates} 
\begin{table*}
\centering
\caption{Values calculated by using PBE for an adatom (A) adsorbed to each $(4 \times 4)$ supercell of the GaS substrate: Adsorption site; 
the adsorption energy $E_{ads}$; the height (distance) of the adatom from the original, high-lying S atomic plane of the substrate $h$; the 
smallest distance between the adatom and S atom of the substrate $d_{A-S}$; the smallest distance between the adatom and Ga atom of the substrate $d_{A-Ga}$; 
the local magnetic moment $\mu$ (NM for nonmagnetic system); the charge transfer between the adatom and substrate $\Delta \rho$ with positive sign indicating the 
donation of electrons to the substrate; and calculated workfunctions $\phi$.}
\begin{tabular}{ccccccccc}
\hline
Adatom (A) & Site& E$_{ads}$ (eV) & h(\AA) &d$_{A-S}$(\AA)&d$_{A-Ga}$(\AA)& $\mu$ ($\mu_B$) & $\Delta \rho$ (e) & $\phi$ (eV) \tabularnewline \hline
Li         & H   &  1.52           & 1.26   & 2.45          & 3.25          & NM              & 0.99          & 3.50 \tabularnewline
Na         & H   &  1.50           & 1.84   & 2.80          & 3.67          & NM              & 0.99          & 3.32 \tabularnewline
K          & H   &  1.67           & 2.42   & 3.18          & 4.11          & NM              & 0.97          & 3.15 \tabularnewline
Be         & H   &  0.47           & 0.85   & 2.07          & 2.81          & NM              & 1.97          & 4.98 \tabularnewline
Mg         & H   &  0.24           & 3.30   & 3.28          & 4.20          & NM              & 1.81          & 4.30 \tabularnewline
Ca         & Ga  &  0.92           & 1.09   & 2.70          & 2.73          & NM              & 1.07          & 4.15 \tabularnewline
B          & H   &  1.23           & 0.74   & 1.90          & 2.75          & NM              & 0.49          & 3.95 \tabularnewline
Al         & H   &  1.14           & 1.83   & 2.74          & 3.65          & NM              & 0.75          & 3.70 \tabularnewline
Ga         & H   &  1.49           & 1.90   & 2.81          & 3.71          & NM              & 2.27          & 3.57 \tabularnewline
C          & S   &  1.86           & 1.68   & 1.73          & 3.53          & 2.0             & 0.56          & 5.21 \tabularnewline
Si         & S   &  1.22           & 2.18   & 2.18          & 3.92          & 2.0             & 0.20          & 4.20 \tabularnewline
Ge         & S   &  2.57           & 2.22   & 2.38          & 3.78          & 2.0             & 3.63          & 4.29 \tabularnewline
N          & S   &  1.14           & 1.51   & 1.56          & 3.28          & 1.0             & 0.55          & 5.27 \tabularnewline
P          & S   &  0.54           & 2.03   & 2.06          & 3.59          & 1.0             & 0.44          & 4.89 \tabularnewline
As         & S   &  3.66           & 2.06   & 2.22          & 3.67          & 1.0             & 4.04          & 4.40 \tabularnewline
O          & S   &  6.93           & 1.42   & 1.51          & 3.32          & NM              & 1.67          & 6.20 \tabularnewline
S          & S   &  1.94           & 1.97   & 1.97          & 3.69          & NM              & -0.19         & 5.58  \tabularnewline
Se         & S   &  1.41           & 2.01   & 2.14          & 3.82          & NM              & 0.93          & 5.43 \tabularnewline         
\hline \hline
\label{tableS}
\end{tabular}
\end{table*}

Here we examine  the  adsorption of adatoms on GaS and GaSe substrates. 
We considered the following possible adsorption sites which are depicted in Fig.~\ref{f1} : hollow (H) site is above the center of hexagon unit,
bridge (B) site is above the midpoint of Ga-X bond, top sites of Ga and X (S or Se) atom is the directly above the Ga or X atoms. Our results are listed in Table \ref{tableS} for the adatoms adsorbed to GaS substrate and in Table \ref{tableSe} for the adatoms adsorbed to GaSe substrate. 
In the following sections, we investigate adsorption mechanism of each adatom in detail.

\begin{table*}
\centering
\caption{Values calculated by using PBE for an adatom (A) adsorbed to each $(4 \times 4)$ supercell of the GaSe substrate: 
Adsorption site of equilibrium structure; the adsorption energy $E_{ads}$; the height (distance) of the adatom from the original, high-lying Se atomic plane of 
the substrate $h$; the smallest distance between the adatom and Se atom of the substrate $d_{A-Se}$; the smallest distance between the adatom 
and Ga atom of the substrate $d_{A-Ga}$; the local magnetic moment $\mu$ (NM for nonmagnetic system); the charge transfer between the adatom 
and substrate $\Delta \rho$ with positive sign indicating the donation of electrons to the substrate; calculated work functions $\phi$.}
\begin{tabular}{ccccccccc}
\hline
Adatom (A) & Site& E$_{ads}$ (eV) & h(\AA) &d$_{A-Se}$(\AA)&d$_{A-Ga}$(\AA)& $\mu$ ($\mu_B$) & $\Delta \rho$ (e)& $\phi$ (eV) \tabularnewline \hline
Li         & H   &  1.36           & 5.88   & 2.59          & 3.43          & NM              & 0.99         & 3.44 \tabularnewline
Na         & H   &  1.43           & 1.98   & 2.95          & 3.88          & NM              & 0.99          & 3.24 \tabularnewline
K          & Ga  &  1.85           & 2.49   & 3.31          & 3.71          & NM              & 0.61          & 2.98 \tabularnewline
Be         & Ga  &  1.47           & -0.01  & 2.24          & 2.13          & NM              & 0.29          & 4.99 \tabularnewline
Mg         & H   &  0.32           & 2.53   & 3.33          & 4.29          & NM              & 1.86          & 4.27 \tabularnewline
Ca         & Ga  &  1.23           & 1.11   & 2.81          & 2.75          & NM              & 0.97          & 4.05  \tabularnewline
B          & H   &  1.22           & 0.89   & 2.15          & 2.97          & NM              & 0.29          & 3.98 \tabularnewline
Al         & H   &  1.12           & 1.90   &2.86           & 3.81          & NM              & 0.56          & 3.53 \tabularnewline
Ga         & H   &  1.49           & 1.96   & 2.91          & 3.87          & NM              & 1.19          & 3.50 \tabularnewline
C          & Se  &  1.59           & 1.76   & 1.93          & 3.72          & 2.0             & 0.46          & 4.90 \tabularnewline
Si         & Se  &  1.20           & 2.33   &2.33           & 4.08          & 2.0             & 0.19          & 4.02 \tabularnewline
Ge         & Se   &  2.60           & 2.20   & 2.52          & 3.78          & 2.0             & 3.45          & 4.12 \tabularnewline
N          & Se  &  0.60           & 1.50   & 1.76          & 3.33          & 1.0             & 2.99          & 4.59 \tabularnewline
P          & Se   & 0.37               & 2.21       & 2.21              & 3.95              & 1.0             & 0.03              &4.66      \tabularnewline
As         & Se   &  3.69           & 2.02   & 2.36          & 3.71          & 1.0             & 4.35          & 4.23 \tabularnewline
O          & Se  &  6.32           & 1.43   & 1.69          & 3.52          & NM              & 1.63          & 5.64 \tabularnewline
S          & Se  &  1.86           & 2.12   &2.12           & 3.85          & NM              & -0.30         & 5.26 \tabularnewline
Se         & Se  & 1.42           & 1.97   & 2.28          & 3.96          & NM              & -0.07         & 5.10 \tabularnewline         
\hline \hline
\label{tableSe}
\end{tabular}
\end{table*}

\subsection{Optimized Structures and energetics of adatoms}

\begin{figure}
\includegraphics[scale=0.4]{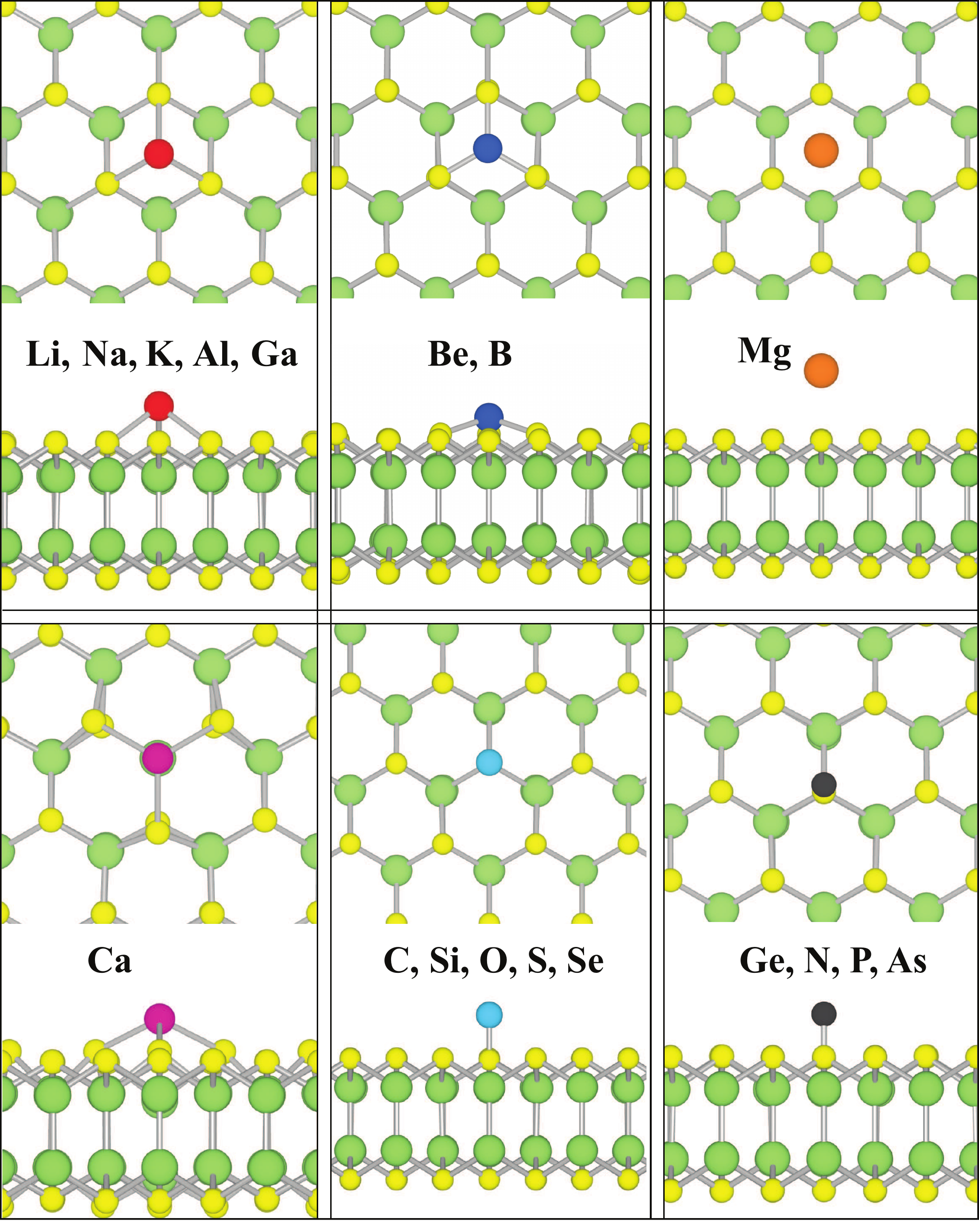}
\caption{(Color online) The optimized atomic structure of the adsorbed adatoms on GaS substrate with both top and side views. (green:~Ga, yellow:~S atoms and, adsorbed adatoms 
are indicated 
by balls with different colors and sizes.}
\label{f2}
\end{figure}

\begin{figure}
\includegraphics[scale=0.40]{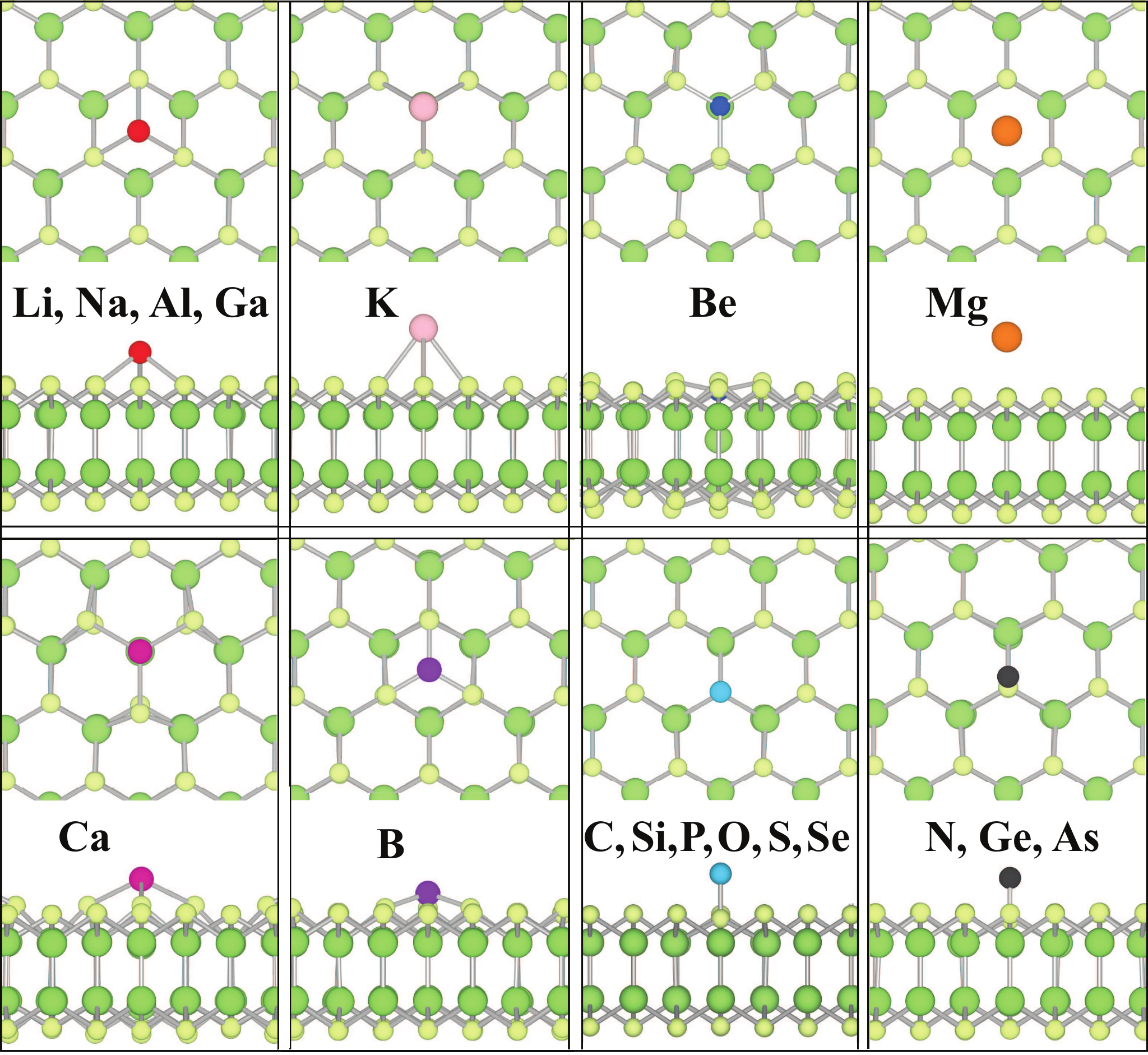}
\caption{(Color online) The optimized atomic structure of the adsorbed adatoms on GaSe substrate with both top and side views. (green:~Ga, dark-green:~Se atoms and, adsorbed adatoms 
are indicated by balls with different colors and sizes.}
\label{f3}
\end{figure}

In order to explore equilibrium structures of adatom adsorbed to GaX systems, adatoms were fully relaxed on $(4 \times 4)$ hexagonal GaX supercell and energetically most stable conformations were obtained as shown in 
Fig.~\ref{f2} and ~\ref{f3}. The adsorption energies of adatoms such as in the same group of Periodic Table are mostly in the same range except for O adatom. Also, adatoms from the same group of Periodic Table are adsorbed at the same adsorption site.
Adsortion energies and structural parameters are listed in Table I and II.
The adsorption energy values show that all these adatoms can chemically adsorb on GaS or GaSe sheets and strongly bind to the monolayers except for Mg atom. Mg shows lowest E$_{ads}$ values on GaX monolayers however it is not surprising due to weak bonding nature of Mg atom with early 3d transition metals \cite{shin}. However, O atom shows the highest E$_{ads}$ value on GaX monolayers and it is followed by As atom. Alkali metal atoms also shows high E$_{ads}$ values relatively.

The binding  energy for the H site, illustrated in Fig.~\ref{f2}, is found to be minimum for  Li, Na, K, Al, Ga, Be, B and Mg atoms adsorbed to GaS monolayer. 
Geometric structure of B and Be atom is shown separately from the figure of Li,Na,K,Al and Ga adsorption, because B and Be are much closer than which of those and similarly Mg atom yields high distance compare with other H site adsorption figures therefore we figured that in a different column.
Among 2A group adatom, Ca prefers bind to Ga top site which is different from Be and Mg atoms which prefer bind over H site. Since Ca atom pushes the closer S atom outward and Ga atom downward direction, Ca deforms the hexagon ring of GaX monolayers.
While C, Si, O, S and Se atoms attach to the substrate onto S site directly, Ge, N, P and As atoms do not bind to S site exactly vertical, in other words they a little bit 
oriented to the B site but much more closer to the S site.
Moreover they do not cause any significant deformation which is shown in Fig.~\ref{f2} separately. 

It is also realized that the adatoms in the third period of periodic table have smaller adsorption energies than second and fourth period of adatoms except for S adatom. 
For example the third period of Si adatom adsorption energy (1.22~eV) is lower than the second and fourth period of adatoms C (1.86~eV) and Ge (2.57~eV), respectively.

Adsorption geometries of GaSe substrates are illustrated in Fig.~\ref{f3}.
Most of adatoms prefer same adsorption site of which GaS monolayer except for K and Be atoms.
While only the Ca adatom prefers to adsorb at Ga site on GaS 
substrate, both Ca and K adatoms prefer to adsorb at top Ga site on GaSe substrate among the chosen atoms in this study. Similar situations are obtained for adatoms on GaSe substrates, such as the same group elements preserve the 
same adsorption sites and the adsorption energy of second period of periodic table smaller than the third and fourth period except for S atoms. 
 1A, 3A groups of adatoms are adsorbed on H site, while 4A, 5A and 6A groups of adatoms are adsorbed on Se site. 
Apart from all these, as in the case of adatoms on GaS substrates, all adatoms stayed above the upper layer of GaSe sheet except for Be. As an exception, the Be adatom by itself
is implemented into the GaSe sheets by pushing the nearest Se atoms upper planes and one Ga atom inside so forms strong bonds with substrate atoms. 

\subsection{Electronic and Magnetic Properties}
In this section, we pay main attention to the electronic and magnetic properties of adatom+substrate systems for their stable structures since adsorption of adatoms can modify the electronic and magnetic properties of GaS and GaSe structures.  
In general, adatom adsorption on substrate is expected to alter the Fermi level, and there can be a large change in work function relative to isolated structure. We define the work function as;
\begin{equation}
 \centering
 \Phi~=~ E_{vac}~-~E_{F} \nonumber
 \end{equation}

 where $\Phi$ defined as the energy required to extract an electron from Fermi level to the vacuum level, E$_{vac}$ is determined the vacuum energy 
 from the adatom+substrate system in the \textit{z}-direction and E$_F$ is the Fermi energy of the system. 
 
Work function $\Phi$ of bare GaS and GaSe supercells are obtained as 6.17 eV and 5.76 eV respectively. This result is reasonable since the linear relationship between work function and ionization energy (S atom has higher ionization energy than Se atom). Results of adsorbed systems are listed in Table \ref{tableS} and \ref{tableSe}, and these results are useful to indicate the trend in $\Phi$ for different adsorbed atoms systematically. As expected, work function for each adatom on GaX substrates are directly proportional with the ionization energy. When nonmetal adatoms are adsorbed on GaX substrate, $\Phi$ does not change so much compared with the bare GaX systems, however in the case of alkali and poor metals, also alkali earth metals, $\Phi$ values decrease compared with the bare GaX structures. 
  
\begin{figure*}
\centering
\includegraphics[scale=0.35]{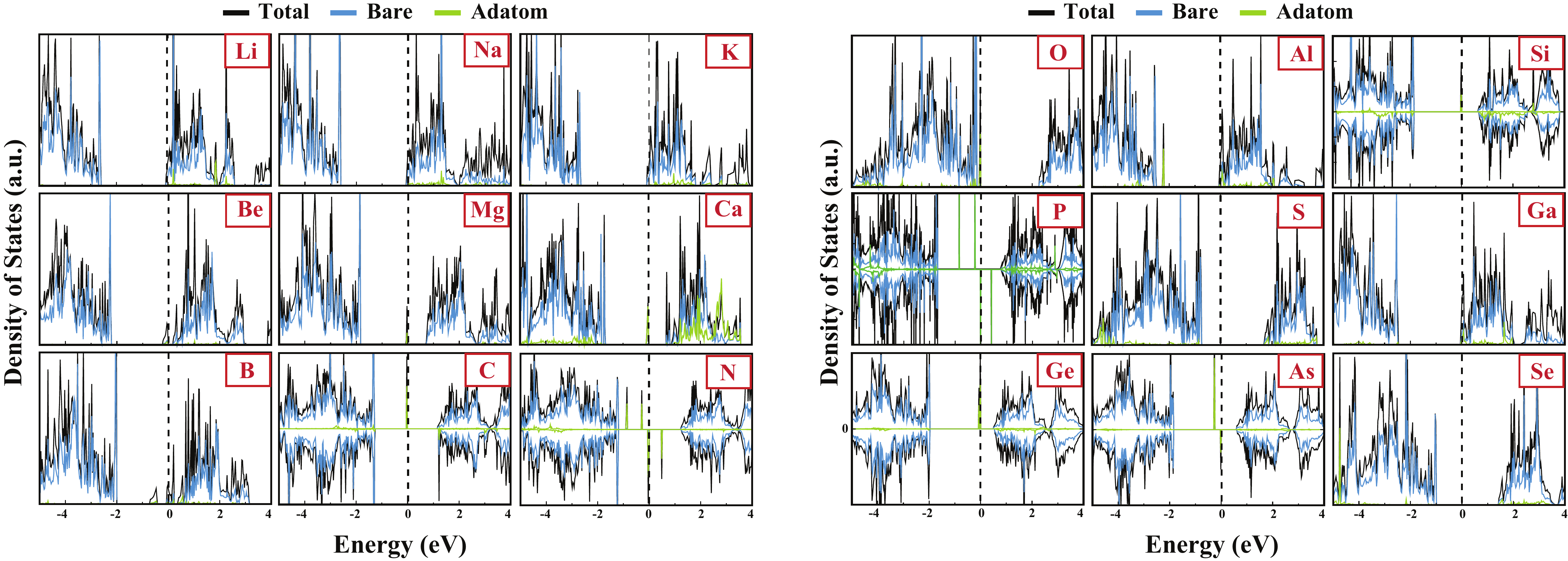}
\caption{(Color online) Calculated density of states for an adatom adsorbed to each $(4 \times 4)$ supercell of GaS at optimized (equilibrium) sites. Total density of states (TDOS) 
is shown by black lines and green line indicate adatom projected density of states. The density of states of the extended GaS substrate is shown by blue tone; which is obtained 
from the local density of states calculated at hosts Ga and S atoms farthest from the adatom. The zero of
the energy is set at the common Fermi level shown by dashed vertical line.}
\label{f4}
\end{figure*}

\begin{figure*}
\centering
\includegraphics[scale=0.35]{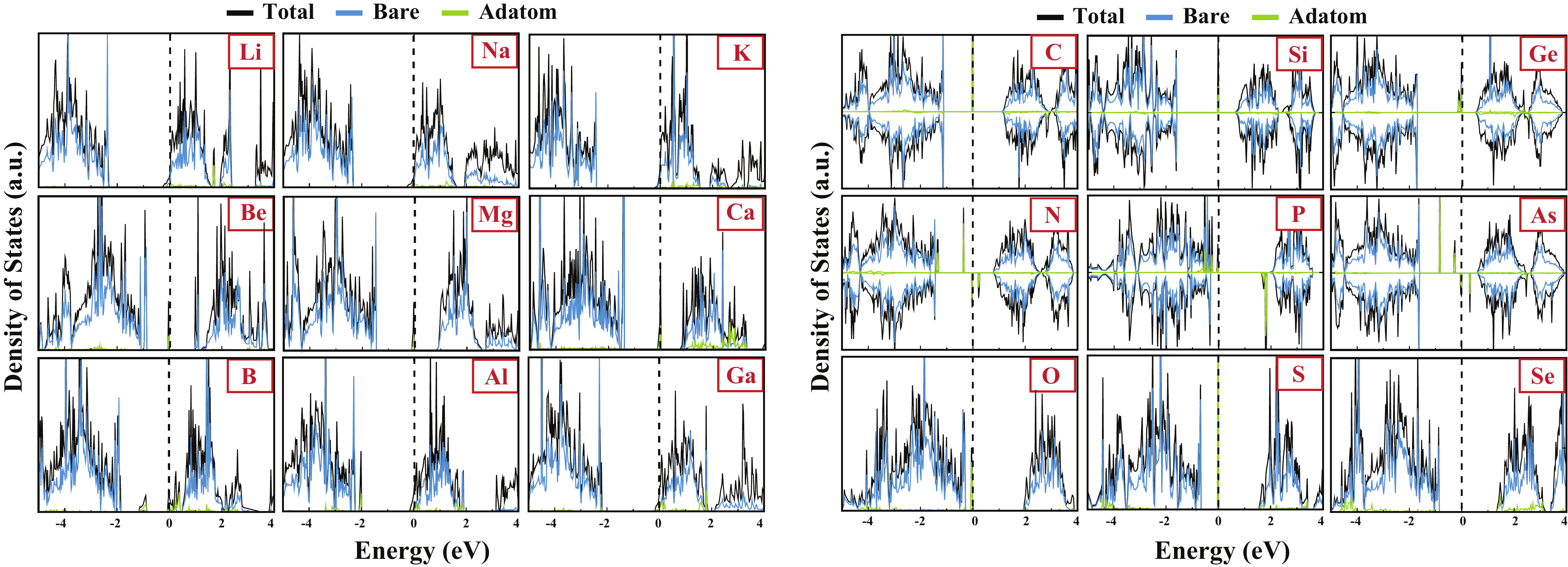}
\caption{(Color online) Calculated density of states for an adatom adsorbed to each $(4 \times 4)$ supercell of GaSe at optimized (equilibrium) site. Total density of states (TDOS) 
is shown by black lines and green line indicate adatom projected density of states. The density of states of the extended GaSe substrate is shown by blue tone; which is obtained 
from the local density of states calculated at hosts Ga and Se atoms farthest from the adatom. The zero of
the energy is set at the common Fermi level shown by dashed vertical line.}
\label{f5}
\end{figure*}

We observed that magnetism can be induced in GaS and GaSe sheets by only the group 4A and 5A non-metallic atoms (also by poor-metal Ge) adsorption. The C, Si, Ge, N, P, As 
adsorbed GaS and GaSe sheets favor the spin-polarized states as seen from the results given in Table 1 and 2. Group 4A adatom C, Si, Ge adsorbed systems have larger magnetic 
moment of 2~$\mu_B$ than group 5A adatoms N, P, As adsorbed systems which have a value of  1~$\mu_B$ for both GaS and GaSe sheets. 

Density of states (DOS)  of each adatom adsorbed to GaS and GaSe monolayer on the favored adsorption sites are presented in Fig.~\ref{f4} and \ref{f5}, respectively. 
The present analysis uses a scheme to deduce the effects of adsorption, such that both figures included the DOS of the "bare" extended substrate, the total and projected densities of states (PDOS) of adatom to find out the energy shifts of adatom localized states relative to the "bare" extended substrate. 

The DOS related with the single adatom adsorbed to GaS substrate are illustrated in Fig.~\ref{f4}. We start to DOS analysis with alkali metal adsorption (Li, Na, K) on GaS 
substrate. These adatoms donate a part of their valance charge to the states which overlaps with the 
bottom of the conduction band of the extended substrate. However, 
both minimum of the conduction and maximum of the valance band are shifted downward direction and all of the structures are still large gap semiconductor. Unlike the 
elements group 1A alkali metals, alkali earth metals (Be, Mg and Ca) significantly alter the GaS electronic structure. An important common feature in the DOS for these adatoms 
on GaS is the hybridization of the adatom and GaS states which is evident from prominent peaks in the DOS as shown in Fig.~\ref{f4}. The impurity states occur inside the 
gap of extended substrate, so the gap is reduced compared to bare structure. Adatoms from group 3A display similar situation with group 1A adatoms except for B atom, since the electronic structure of B adsorbed to GaS 
sheet turns to metallic. The impurity states occur very close to the Fermi level and cross each other near to the Fermi level. 
States derived from C, Si, and Ge adatoms adsorbed to GaS are spin-polarized and appear near the Fermi level. Spin-up states occur below the Fermi level and hence become 
occupied. Similar characteristics for the DOS are found for group 5A adatoms (N, P, As), however energy levels of both these group adatoms are splitted to spin-up and spin-down channels resulting a
net magnetic moment. For the case N adsorption to the GaS substrate, two spin-up impurity states are occupied at the -0.36 and -0.92 eV below the Fermi level, while in the 
spin-down channel, one of the impurity state is occupied at the -0.1 eV below the Fermi level and the other one is empty at the 0.42 eV above the Fermi level. P and As adatom display 
similar situation like N adatom, therefore the fundamental band gap of structure is locally reduced for these adatoms. We now discuss the DOS for O adatom which is representative of 
group 6A adatoms S and Se. O derived states occur in the valance band, however, S and Se adatom derived states occur in conduction band but low in intensity.
Nevertheless, these adatoms adsorbed to GaS substrate preserve the fundamental wide band gap as semiconducting in electronic property.  
The DOS figures related with single adatom adsorbed to GaSe substrate are illustrated in Fig.~\ref{f5}. Comparing with Fig.~\ref{f5} with Fig.~\ref{f4}, trends mentioned
above are similar with corresponding adatoms with insignificant changes in electronic states. 

The amount of charge accumulation is different for most of adatoms treated on GaS and GaSe sheets which are given in Table~1 and 2. 
Bader analysis results give that all adatoms except for S atom (also Se atom on GaSe sheet, but it is too small, so ignorable), donate charge to GaS and GaSe substrate.  
We assigned a negative sign to the calculated value of $\Delta \rho$ if charge 
is transferred from substrate to adatom. S atom on GaX sheets take small amount 
of charges from substrates. Se atom gives small amount of charge to S atom on GaS substrate because of higher electronegativity of S atom compare with Se atom, relatively. 
The most adatom is stabilized closer to the surface, thus ionic bonding appears and overall hexagonal geometry has been retained.

\section{Conclusions}\label{conc}
In this paper, adsorption of 18 different adatoms on GaX is studied using the first principles density functional theory. Calculations of adsorption energy, 
geometric structure, density of 
states, charge transfer and work function give a consistent picture for the adatoms considered. 
Results show that all adatoms considered in this study chemically adsorb to GaX monolayers except for Mg atom due to weak bonding nature of Mg atom with early 
3d transition metals, however O atom shows the highest E$_{ads}$ value on GaX monolayers. 
Adatoms from groups I-III on GaS substrate exhibit characteristic of ionic bonding 
(except for Mg) and very small distortion of the GaX sheet with a little changes in the GaX electronic states. Significant charge transfers are observed except 
for S and Se atoms and total magnetic moments are found by adsorption of adatoms which is essential for spintronic applications. Work function results provide useful 
knowledge of trend in $\Phi$ for different adsorbed atoms systematically. Our study represents adsorption strategies of GaX monolayers therefore a theoretical knowledge 
for future experimental and theoretical studies.


\begin{acknowledgments}

We would like thank to TUBITAK ULAKBIM, High Performance and Grid Computing Center for numerical calculations.
H. D. O. and E. A. acknowledge support from TUBITAK through the project No. 116F059.
\end{acknowledgments}

\bibliography{rsc3}

\end{document}